\documentclass[aps,
 amsmath,amssymb,
prd,
twocolumn
]{revtex4}

\usepackage{graphicx}
\usepackage{dcolumn}
\usepackage{bm}
\usepackage[utf8]{inputenc}
\usepackage{physics}
\usepackage[colorlinks=true,linktoc=all]{hyperref}
\usepackage{xcolor}
\usepackage{changebar}
\usepackage{mathtools}

\newcommand{\RH}{\mathrm{RH}}

\newcommand{\lsim}   {\mathrel{\mathop{\kern 0pt \rlap
  {\raise.2ex\hbox{$<$}}}
  \lower.9ex\hbox{\kern-.190em $\sim$}}}
\newcommand{\gsim}   {\mathrel{\mathop{\kern 0pt \rlap
  {\raise.2ex\hbox{$>$}}}
  \lower.9ex\hbox{\kern-.190em $\sim$}}}

\begin{document}

\title{Gravitational leptogenesis from metric perturbations}

\author{Antonio L.\ Maroto}
\email{maroto@ucm.es}
\affiliation{Departamento de F\'{\i}sica Te\'orica and Instituto de F\'{\i}sica de Part\'{\i}culas y del Cosmos (IPARCOS-UCM), Universidad Complutense de Madrid, 28040 
Madrid, Spain}

\author{Alfredo D. Miravet}
\email{alfrdelg@ucm.es}
\affiliation{Departamento de F\'{\i}sica Te\'orica and Instituto de F\'{\i}sica de Part\'{\i}culas y del Cosmos (IPARCOS-UCM), Universidad Complutense de Madrid, 28040 
Madrid, Spain}



\begin{abstract} 
In this work we make the observation that the gravitational leptogenesis mechanism can be implemented without invoking new axial couplings in the inflaton sector. We show that in the perturbed Robertson-Walker background  emerging after inflation, the spacetime metric itself breaks parity symmetry and generates a non-vansihing Pontryagin density which can produce a matter-antimatter asymmetry. We analyze the produced asymmetry in different inflationary and reheating scenarios.
We show that the generated asymmetry can be locally comparable to observations in certain cases, although the size of the matter-antimatter regions is typically  much smaller than the present Hubble radius.
\end{abstract} 

\maketitle

\section{Introduction}
The excess of matter over antimatter in the universe is one of the 
longstanding problems in cosmology \cite{Sakharov:1967dj}. This matter-antimatter 
asymmetry is usually quantified through the ratio of the net
baryon number density with respect to the total entropy density 
whose value measured by the Planck collaboration is $n_B/s=8.718\pm 0.004\times 10^{-11}$ \cite{Planck:2018vyg}. One 
of the most interesting proposals for the generation of the baryon asymmetry is leptogenesis \cite{Fukugita:1986hr}. The original implementation of this mechanism relied on the introduction of right-handed Majorana neutrinos in the Standard Model, whose mass term breaks lepton symmetry. This 
lepton asymmetry is later on converted into baryon asymmetry through electroweak sphaleron processes \cite{Kuzmin:1985mm}. For right-handed neutrinos in thermal equilibrium, this mechanism requires a reheating temperature $T_{\RH}$ above the right-handed neutrino mass $m_R$ which should satisfy  $m_R\gsim 10^9$ GeV 
\cite{Buchmuller:2004nz,Kamada:2019ewe}. For non-thermally produced neutrinos these constraints could be relaxed  \cite{Pilaftsis:2003gt,Co:2022bgh, Asaka:1999yd}.

In \cite{Alexander:2004us}, an alternative mechanism for leptogenesis was
proposed which is not based on the introduction of heavy Majorana leptons. In this gravitational leptogenesis mechanism, lepton asymmetry is generated from
the chiral gravitational lepton anomaly already operating in the Standard Model with only left-handed neutrinos \cite{Alvarez-Gaume:1983ihn,delRio:2021bnl}.
\begin{eqnarray} \label{eq:JL_div}
\nabla_\mu J^\mu_L=\frac{N_{R-L}}{24(4\pi)^2}R\tilde R
\label{anomaly}
\end{eqnarray}
where $J^\mu_L$ is the total lepton current and  $N_{R-L}$ is the difference between the number of right-handed and left-handed lepton species. As a matter of fact, it has been shown that neutrino masses, either Dirac or Majorana, do not affect the predictions of the gravitational leptogenesis \cite{Adshead:2017znw}.

 The necessary ingredient in this case for the generation
 of a net lepton number is the existence of a primordial chiral gravitational wave background 
which contributes to the Pontryagin density $R\tilde R$. In order to generate such a chiral background, extended inflationary models involving axial couplings of the 
inflaton field have been considered. Thus for example, a gravitational Chern-Simon coupling of a pseudo-scalar inflaton field was originally proposed in \cite{Alexander:2004us}, although some consistency issues were discussed in \cite{Alexander:2004wk, Lyth:2005jf}.
Other possibilities include a Chern-Simons 
interaction between the
pseudo-scalar inflaton and a $U(1)$ gauge field \cite{Papageorgiou:2017yup} and non-abelian gauge fields coupled to an axionic inflaton \cite{Maleknejad:2012wqk,Caldwell:2017chz}. Alternative ways of generating a parity-violating GW background have been considered in \cite{Kawai:2017kqt, Abedi:2018top, Barrie:2021orn}.

In this work we make the observation that the gravitational leptogenesis mechanism can be implemented without invoking  new  axial couplings in the inflaton sector. Indeed, the perturbed Robertson-Walker (RW) background  emerging after inflation already breaks parity thus 
generating a non-vansihing Pontryagin density. Notice that 
although the probability distribution functions for the production of  
left and right handed gravity waves are the same in ordinary inflation models, our universe is a particular realization of the Gaussian 
process in which the actual amplitude of left and right handed  gravitational 
wave excitations can be different.

We thus conclude that the  minimal  Standard Model with left-handed neutrinos together with an ordinary  inflationary model driven by a scalar inflaton field already contains all the ingredients to generate a lepton asymmetry after inflation. 

\section{Gravitational leptogenesis}
Let us then consider a spatially flat RW spacetime with scalar and tensor perturbations in the longitudinal gauge. We will ignore vector perturbations as they are not typically produced during inflation. The line element in conformal time reads
\begin{equation}
    \dd{s}^2 = a^2(\eta)\left[(1+2\Phi)\dd{\eta}^2 - ((1-2\Psi)\delta_{ij} - h_{ij}) \dd{x}^i \dd{x}^j\right], \label{FLRW}
\end{equation}
with $\Phi, \Psi$ the scalar perturbations and $h_{ij}$ the transverse traceless tensor perturbation.

The Pontryagin density that sources the leptonic current in \eqref{anomaly} can be written in terms of the electric and magnetic parts of the Weyl tensor \cite{delRio:2020cmv, Hwang:1990am, Goode:1989jt}, ${E}$ and ${B}$ respectively, as
\begin{equation}
    R\tilde R = \frac{1}{2} \varepsilon^{\mu \nu} {}_{\rho\sigma} R_{\mu\nu\alpha\beta} R^{\rho\sigma\alpha\beta} = 16 E_{\mu\nu} B^{\mu \nu}.
\end{equation}

\begin{figure*}
    \centering
    \includegraphics{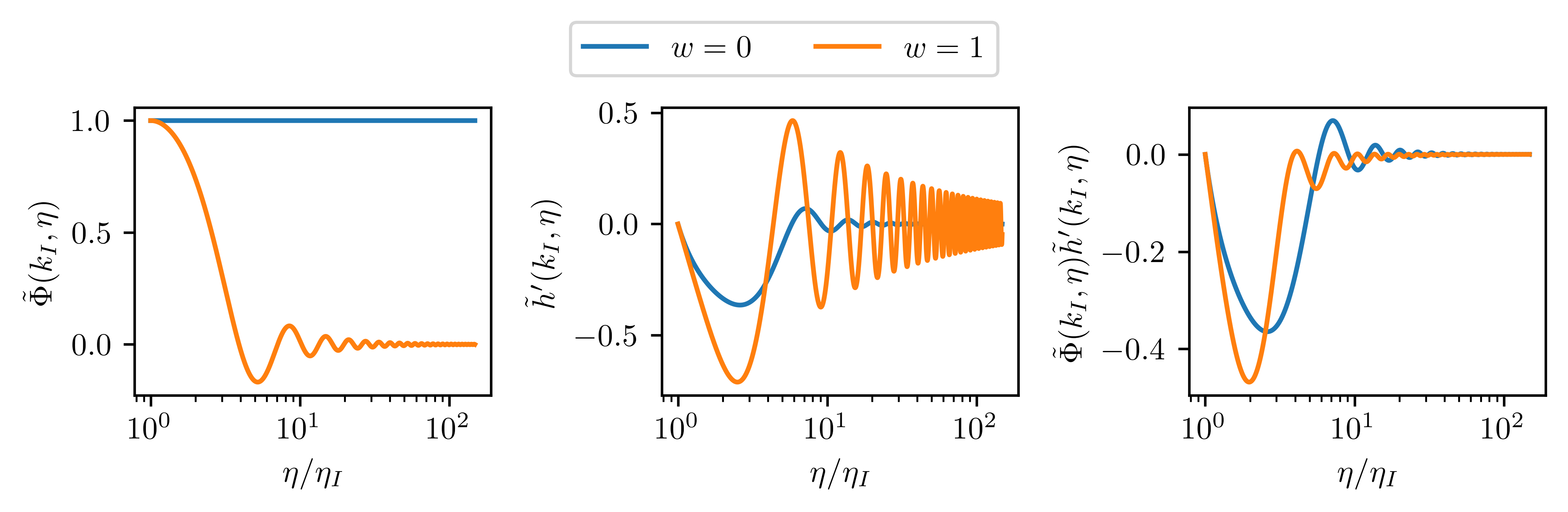}
    \caption{Time evolution of the perturbations $\tilde\Phi(k_I,\eta)$ (left), $\tilde{h}'(k_I,\eta)$ (center) and its product (right), which is the integrand of the time integral in Eq. \eqref{eq:I}, for  reheating equations of state $w=0,1$ and wavenumber $k=p=k_I$.}
    \label{fig:integrand}
\end{figure*}

The unperturbed part of the metric \eqref{FLRW} does not contribute to either the electric or magnetic parts of the Weyl tensor. On the other hand,  all scalar, vector and tensor components contribute to the electric part, whereas only vector and tensor ones add to the magnetic part. This means that the leading contribution to $R\tilde R$ is second order in metric perturbations. Since the tensor-to-scalar ratio of the primordial power spectra $r< 0.1$ \cite{Planck:2018jri}, we expect  the scalar-tensor contribution to dominate over the tensor-tensor one, i.e.
\begin{equation}\label{eq:RR_scalartensor}
    R \tilde R = -\frac{4}{a^4}\epsilon_{jkl}(\Phi + \Psi)_{,ij} h_{ik,l}' + \dots
\end{equation}
where prime denotes derivative with respect to the conformal time. In the comoving frame we can write $J^\mu_L=(a^{-1}n_L,0)$ to leading order in perturbations, where $n_L$ is the physical lepton number density. Inserting these expressions into \eqref{eq:JL_div} we obtain the leptonic number density after integrating in time
\begin{equation}
    n_L = \frac{1}{16\pi^2 a^3}\epsilon_{jkl} \int\dd{\eta} \Phi_{,ij} h_{ik,l}',
\end{equation}
where we used that in the Standard Model $N_{R-L}=-3$, and that in the absence of anisotropic stress $\Phi = \Psi$.

Let us now expand the scalar and tensor perturbations in terms of creation and annihilation operators 
\begin{align}
&\Phi(\eta,\vb x)=\int \frac{\dd[3]{p}}{(2\pi)^{3/2}} \left(\Phi(p,\eta)b_{\vb p}\,e^{i\vb p\cdot \vb x}+\Phi^*(p,\eta)b^\dagger_{\vb p}\,e^{-i\vb p\cdot \vb x}\right)  
\end{align}
and 
\begin{align}
&h_{ij}(\eta,\vb x)=\int \frac{\dd[3]{k}}{(2\pi)^{3/2}}  \sum_{\lambda=+, -}\left(h_\lambda(k,\eta)e_{ij}^\lambda(\vb{\hat k})a_{\vb k,\lambda}e^{i\vb k\cdot \vb x}\right.\nonumber \\
&\left.+h_\lambda^*(k,\eta)e^\lambda_{ij}{}^*(\vb{\hat k})a^\dagger_{\vb k,\lambda}e^{-i\vb k\cdot \vb x}\right),
\end{align}
where $\lambda=\pm$ correspond to the $\pm 2$ helicity modes whose polarization tensors can be written as $e_{ij}^\lambda(\vb{\hat k}) = \varepsilon_i^{\lambda *} (\vb{\hat k}) \varepsilon_j^{\lambda *} (\vb{\hat k})$, with $\pmb{\varepsilon}^\lambda (\vb{\hat k})$ the helicity $\pm 1$ polarization vectors. Notice that in the absence of chiral couplings 
$h_+=h_-=h$.

It is straightforward to see that the expectation value of the scalar-tensor contribution to the lepton number in the Bunch-Davies vacuum
is vanishing.  Indeed, we can schematically write
$\langle n_L \rangle \sim \langle\Phi h'\rangle = \langle\Phi\rangle \langle h' \rangle = 0$. Therefore, the leading contribution to the expectation value would be the tensor-tensor one, which has been already explored in previous works \cite{Alexander:2004us,Papageorgiou:2017yup, Kamada:2019ewe}. However this contribution also vanishes for inflationary sectors without axial couplings.
Notice however that the variance of the lepton number density is in general non-vanishing since we can write 
$\langle n_L^2 \rangle \sim \langle\Phi^2\rangle \langle h'^2 \rangle \neq 0$. Precisely the root mean square $n_L^{\text{rms}}= \langle n_L^2 \rangle^{1/2}$ provides an estimate of the produced lepton density in a typical
realization of the random process. Thus, it is straightforward to obtain
\begin{align}\label{eq:RRvar}
    &\langle n_L^2 \rangle = \left(\frac{1}{16\pi^2 a^3}\right)^2 \sum_{\lambda}  \int\frac{\dd[3]{\vb{k}} \dd[3]{\vb{p}} \dd{\eta} \dd{\eta'}}{(2\pi)^6}  \nonumber\\
    &\times k^2 |\vb{p}\cdot \pmb{\varepsilon}^\lambda(\vb{\hat k})|^4\left( h'_\lambda(k,\eta) h^{*\prime}_{\lambda}(k,\eta') \Phi(p,\eta) \Phi^*(p,\eta') \right)
\end{align}

In the absence of chiral couplings the whole integral in \eqref{eq:RRvar} is independent of $\lambda$ thanks to spherical symmetry. After some simplification, the variance can be written in a compact manner as

\begin{eqnarray}\label{eq:n2L_general}
    \langle n^2_L \rangle = \frac{1}{3840\pi^8 a^6}\int \dd{k}\dd{p} k^4 p^6 \left|\int \dd{\eta} \Phi(p,\eta) h '(k,\eta)\right|^2 \label{n2}
\end{eqnarray}

\section{Leptogenesis during reheating}
For the sake of concreteness, we will assume that the net lepton number density at the end of inflation is negligible, so that we will consider the leptogenesis produced throughout the stage of reheating by the inflationary primordial metric perturbations. For simplicity, we consider that the energy content during reheating is described by means of an effective fluid with barotropic equation of state $p=w\rho$, with $w$ constant. We will also parametrize the primordial power spectra in the usual way:
\begin{equation}
    \mathcal{P}_S(k) = A_S \left(\frac{k}{k_*}\right)^{n_S-1}, \quad \mathcal{P}_T(k) = A_T \left(\frac{k}{k_*}\right)^{n_T},
\end{equation}
where $A_S$ and $A_T$ are the scalar and tensor amplitudes at the pivot scale $k_*$, respectively. We will use the value obtained by the Planck collaboration \cite{Planck:2018vyg} for the scalar spectral index $n_s = 0.965$ and assume a scale-invariant tensor power spectrum $n_T=0$. The primordial power spectra generated during inflation have a natural ultraviolet cutoff 
at the scale $k_I=a_I H_I$ corresponding to the size of the comoving Hubble horizon at the end of inflation, as modes with $k>k_I$ have never left the horizon and could not become classical metric perturbations. Imposing this upper limit in the momentum integrals in \eqref{n2}, we obtain for the total lepton number variance generated during reheating
\begin{align}\label{eq:nL2_RH}
    \langle n_L^2& \rangle_\RH = \frac{1}{960\pi^4} \left(\frac{k_I}{a_\RH}\right)^6 \mathcal{P}_S(k_I) \mathcal{P}_T(k_I) I(\eta_\RH) 
\end{align}
with
\begin{eqnarray}\label{eq:I}
    I(\eta)&=&\int_0^1 \dd{x} \int_0^1 \dd{y} x^{n_T+1} y^{n_S+2}\nonumber \\
    &\times&\left|\int_{\eta_I}^{\eta} \dd{\hat\eta} \tilde{\Phi}(k_I y,\hat\eta) \tilde{h}'(k_I x,\hat\eta)\right|^2
\end{eqnarray}
where $a_{RH}$ denotes the scale factor at the end
of reheating, $\tilde\Phi(k,\eta) = \Phi (k,\eta) / \Phi(k,\eta_I)$ is the scalar perturbation normalised to its value at the end of inflation, and similarly for the tensor mode. Notice that the
 $x$ and $y$ integrals are dominated by the upper integration limits which correspond to modes with  $k\simeq p\simeq k_I$.  

\begin{figure}
    \centering
    \includegraphics[width=\linewidth]{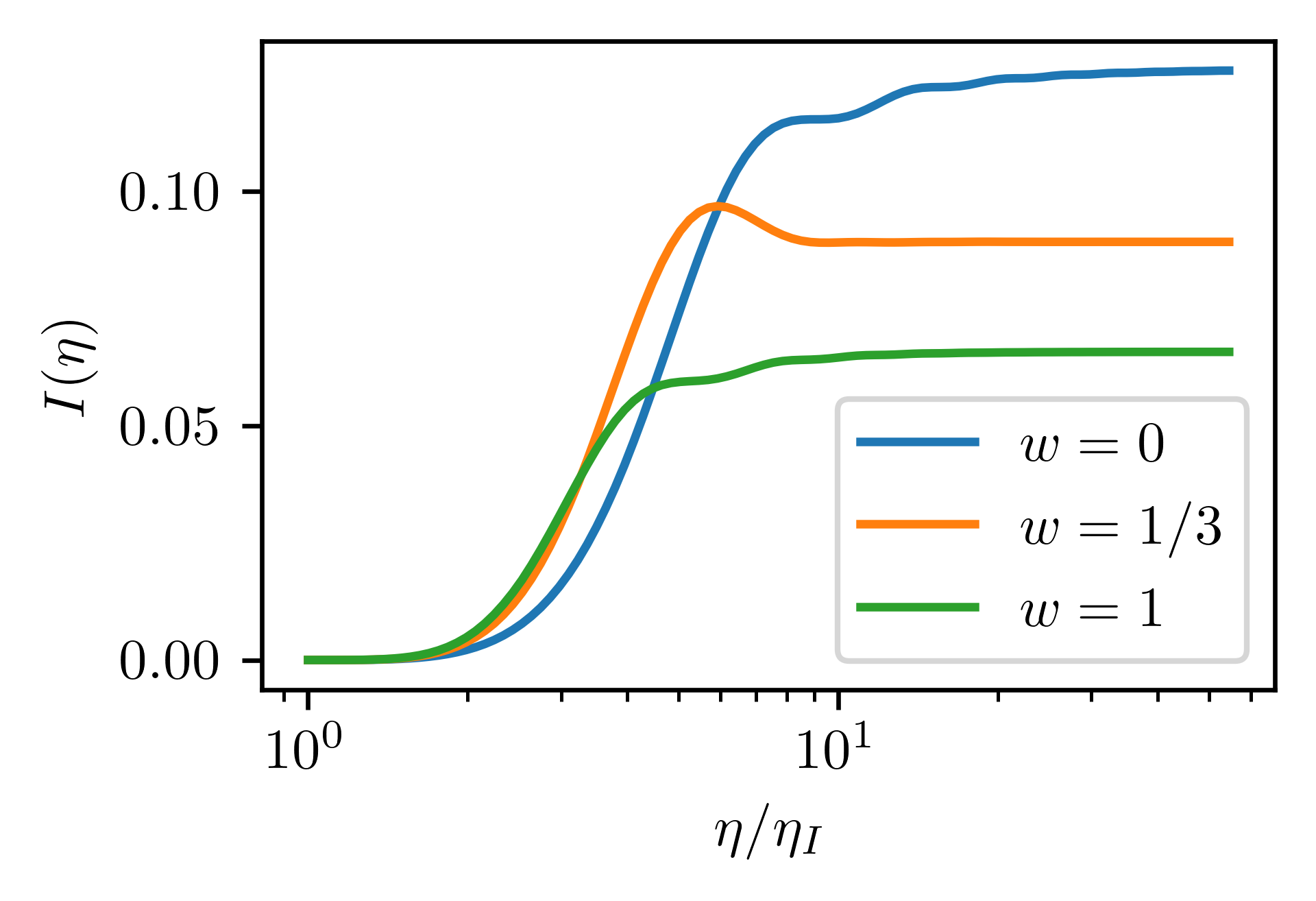}
    \caption{Time evolution of $I(\eta)$ in \eqref{eq:I} for three different reheating equations of state. For inflationary and reheating scales so that $\eta_\RH/\eta_I\gg 1$, the integral evaluates to its asymptotic value and, as a result, it depends only on $w$.}
    \label{fig:integral}
\end{figure}

 During reheating, scalar and tensor modes behave as 
\begin{align}
    \Phi(p,\eta)&=\eta^{-r}[C_1 J_r(\sqrt{w}p\eta)+C_2 Y_r(\sqrt{w}p\eta)]
\end{align} 
and
\begin{align}
h(k,\eta)&=\eta^{s}[D_1 J_s(k\eta)+D_2 Y_s(k\eta)]
\end{align}
with $J_{r,s}$ and $Y_{r,s}$ Bessel functions of order  $r=(5+3w)/(2+6w)$ and $s=(3w-3)/(2+6w)$ respectively and $C_{1,2}$, $D_{1,2}$ constants.  Both scalar and tensor modes remain constant outside the horizon ($k\eta\ll 1$), while well inside the horizon they 
oscillate with decreasing amplitude, except for $w=0$, where $\Phi=\mathrm{const.}$ even inside the horizon. In Fig. \ref{fig:integrand} we show the behaviour of the  perturbations
together with the product appearing in the integrand of \eqref{eq:I} for different values of $w$ for the dominant modes. We can see that contributions to the time integral in \eqref{eq:I} occur only when tensor modes enter the Hubble horizon, since in the super-Hubble regime $h'\simeq 0$.  We also plot function $I(\eta)$ in Fig. \ref{fig:integral} which provides the time dependence of the lepton production. The different behaviour for each equation of state, and hence the different interference between scalar and tensor modes, translates into a different time evolution of $I(\eta)$. We can also see that leptogenesis takes place in a few Hubble times.

We can now compute the lepton number to entropy ratio by dividing by the entropy density $s=2\pi^2 g_{*s} T_\RH^3/45$, with $g_{*s}$ the effective number of relativistic species. This ratio should be so that it reproduces the baryon asymmetry after the partial conversion of leptonic asymmetry via sphalerons \cite{Kuzmin:1985mm}, which is 
\begin{equation}\label{eq:ns_needed}
    \frac{n_L^\mathrm{rms}}{s} = \frac{79}{28}\left|\frac{n_B}{s}\right| = 2.45\times 10^{-10}.
\end{equation}

Far from the instantaneous reheating limit
we find that in general $I(\eta_\RH)=\mathcal{O}(10^{-1})$. Thus, from \eqref{eq:nL2_RH}, we see that, apart from the power spectra which are (almost) scale invariant, the relevant quantity in determining the  size of $n_L^\mathrm{rms}/s$ is the factor  $(k_I/a_\RH)^3$. As a matter of fact, we can extract the dependence on $H_I$ and $T_{\RH}$ from such a factor and obtain
\begin{equation}\label{eq:ns_scaling}
    \left.{\frac{n_L^\mathrm{rms}}{s}}\right|_{\RH} \propto  \sqrt{A_S A_T}\left(\frac{H_I}{M_P}\right)^{\frac{1+3w}{1+w}} \left(\frac{T_{\RH}}{M_P}\right)^{\frac{1-3w}{1+w}}
\end{equation}
for $n_s\simeq 1$ and where $M_P=1/\sqrt{G}$ is the Planck mass, meaning that leptogenesis is enhanced for high inflation scales and, in the case of stiff reheating scenarios with $w>1/3$, for low reheating temperatures.

\begin{figure*}
\centering
\includegraphics[width=\linewidth]{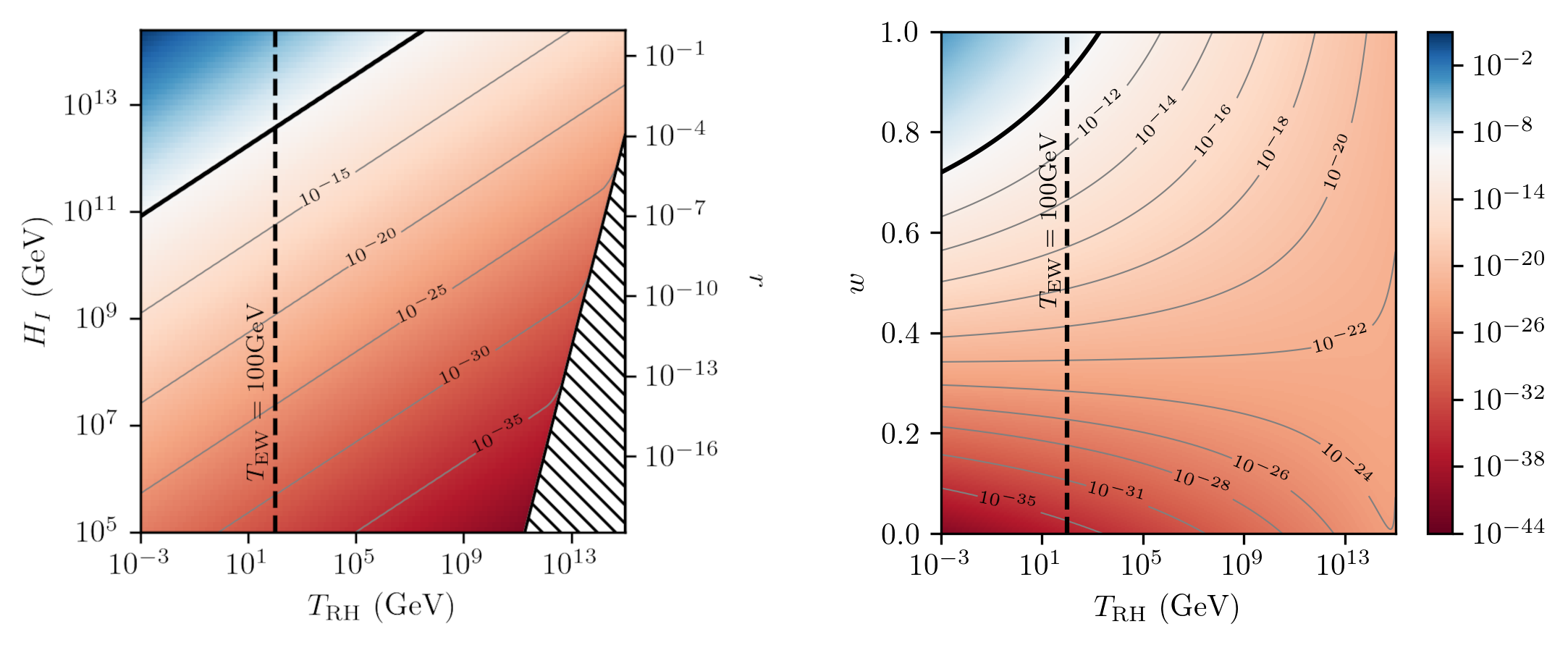}
\caption{Values of $n_L^\mathrm{rms}/s$ for power spectra with Planck 2018 scalar amplitude and scale-invariant tensor power spectrum $A_T=16/\pi(H_I/M_P)^2$ with fixed equation of state $w=1$ (left) and fixed inflation scale $H_I=10^{13}$ GeV (right).  We only consider temperatures $T_\RH > 1$ MeV in order to ensure the 
existence of a Big Bang nucleosynthesis period. The vertical dashed line shows the threshold of the electroweak scale, which is the minimum temperature for the sphaleron process to be effective. The solid black line shows the parameters that yield the value \eqref{eq:ns_needed}, which scales as in \eqref{eq:ns_scaling}. The dashed area on the bottom right corner in the left panel is excluded since reheating is not possible in such a parameter range, with the boundary corresponding to instantaneous reheating. The contour lines (in grey) in the left panel deviate from  \eqref{eq:ns_scaling} close to the limit of instantaneous reheating.}
\label{fig:ns}
\end{figure*}

In the left panel of Fig. \ref{fig:ns}, we plot  $n_L^\mathrm{rms}/s$ in the $(T_\RH, H_I)$ parameter space for a stiff equation of state $w=1$ during reheating. We see that the asymmetry in \eqref{eq:ns_needed} can be locally  generated for inflationary scales above $H_I = 10^{12}$ GeV and reheating temperatures larger than the electroweak threshold.

In the right panel of Fig. \ref{fig:ns} we plot the values of $n_L^\mathrm{rms}/s$ obtained from \eqref{eq:nL2_RH} 
in the $(T_\RH, w)$ parameter space for an inflation scale corresponding to $H_I=10^{13}$ GeV, which corresponds to a tensor-to-scalar ratio of $r\simeq 10^{-3}$. We can see that large baryon asymmetries can be generated for  stiff equations of state and reheating temperatures near the electroweak scale.
For a radiation behaviour $w=1/3$, we find that, as expected, the production is
not sensitive to the reheating temperature. Reheating scenarios with equations of state close to $w=0$ are not  efficient at producing lepton number.
Comparing these results with those obtained in \cite{Alexander:2004us} with the axial coupling, we find that for the instantaneous reheating case and the parameters used in that work, we get $n_L^{rms}/s\sim 10^{-20}$ which is slightly above their results.


\section{Size of matter-antimatter regions}

The variance of the lepton number density obtained in \eqref{eq:nL2_RH} only provides the typical amplitude of local fluctuations. In order to determine the size of the matter-antimatter regions it is necessary to calculate the correlation function
$\xi(\vb{r}) = \langle n_L(\vb{x}+\vb{r}) n_L(\vb{x})\rangle$. For an equation of state during reheating $w=0$ and scale invariant scalar and tensor spectra, it is possible to obtain analytical expressions. Thus, changing variables $\vb{p} = \vb{q}-\vb{k}$, we can write
\begin{align}
    \xi(r) &= \frac{A_S A_T}{4096 \pi^5 a^6} \int_{k_0}^{k_I} \dd{k} \int \dd[3]{\vb{q}} kq^4 e^{i\vb{q}\cdot\vb{r}} \nonumber \\    
    &\times \int_{-1}^1 \dd{x} \frac{(1-x^2)^2}{(q^2+k^2-2kqx)^{3/2}},
\end{align}
where $k_0$ and $k_I$ denote the infrared and ultraviolet cutoffs of the production. The following limits can be obtained
\begin{equation}
    \xi(r) = \begin{cases}
        \dfrac{41 A_S A_T}{483840\pi^4}\left(\dfrac{k_I}{a}\right)^6, & k_I r \ll 1 \\
        -\dfrac{11A_S A_T}{26880\pi^4}\left(\dfrac{k_I}{a}\right)^6 \dfrac{\cos(k_I r)}{k_I^2 r^2}, & k_I r \gg 1
    \end{cases}
\end{equation}

The comoving coherence length associated to the lepton number variance is
therefore $l_\text{coh} \simeq 1/k_I$, which is much smaller than the observable universe. We thus expect that these small matter-antimatter regions will annihilate each other on subhorizon scales, providing an average asymmetry over a  comoving scale $r_0$ given by the weighted variance
\begin{equation}\label{eq:weighed_var}
    \langle n_L^2 \rangle_{r_0} = \left\langle \frac{1}{V_W^2}\left(\int \dd[3]{\vb{r}} n_L(\vb{x}+\vb{r}) W(\vb{r})\right)^2 \right\rangle,
\end{equation}
where $W(\vb{r})$ is a window function of characteristic size $r_0$ and $V_W=\int\dd[3]{\vb{r}} W(\vb{r})$. Again for an equation of state during 
reheating $w=0$ and scale-invariant spectra  we can write
\begin{align}
    \langle n_L^2 \rangle_{r_0} &= \frac{A_S A_T \pi}{256 a^6 r_0^6} \int_{k_0}^{k_I} \dd{k} \int_0^{k_I} \dd{q} kq^6 \frac{|\hat{W}(k+p)|^2}{V_W^2}\nonumber \\
    &\times\int_{-1}^1 \dd{x} \frac{(1-x^2)^2}{(q^2+k^2-2kqx)^{3/2}}.
\end{align}

If we choose a gaussian window function $W(r)=e^{-r^2/2r_0^2}$, we can obtain 
\begin{align}
    \langle n_L^2 \rangle_{r_0} &= \frac{A_S A_T}{2048\pi^4 a^6 r_0^6} \left[ \frac{2e^{-k_0^2 r_0^2}}{k_0^2 r_0^2} + \sqrt{\pi}\left(\left( \frac{1}{k_I^3 r_0^3}-\frac{2}{k_I r_0}\right)\right.\right.\nonumber \\
   &\left.\left. \times\erf(k_I r_0) - \left( \frac{1}{k_0^3 r_0^3}-\frac{2}{k_0 r_0}\right) \erf(k_0 r_0)\right)\right].
\end{align}

 Compared to the local variance in \eqref{eq:RRvar}, this quantity is suppressed as
\begin{equation}\label{eq:nL2_weighed}
    \langle n_L^2 \rangle_{r_0} \simeq \frac{1}{(k_I r_0)^6} \frac{1}{k_0 r_0} \langle n_L^2 \rangle,
\end{equation}
exhibiting a clear blue-tilted behaviour.  
Thus, on regions of order $r_0=H_0^{-1}$, the expected lepton asymmetry will be 
several orders of magnitude smaller than observations. 


\section{Phenomenology in the late universe}

Let us examine the implications of this gravitational leptogenesis mechanism for standard cosmology. Firstly, the lepton number asymmetry generated during reheating can be converted into baryon asymmetry only if $T_\RH$ is above the electroweak scale. Around and above this temperature, QCD confinement has not occurred yet, so the baryon number is in the form of quarks, which are relativistic. Quarks interact in this pre-confinement plasma with a mean free path which can be estimated as $\Gamma_q^{-1}\simeq T^{-1}$, causing diffusion of the baryon number. As a result, perturbations in the baryon asymmetry are suppressed on scales below the corresponding (comoving) Silk length which can be estimated at the moment of confinement as follows \cite{Kolb:1990vq}
\begin{equation}\label{eq:silk_length}
    r_S^2(T_{QCD}) = \int_0^{a_\mathrm{QCD}} \dd{a}\frac{\Gamma_q^{-1}(a)}{a^3 H(a)} \simeq (10^{-16}\text{ Mpc})^2,
\end{equation}
with $a_\mathrm{QCD}$ the scale factor at the confinement temperature $T_\mathrm{QCD}\simeq 300$ MeV.

After confinement, quarks cannot longer exist as free particles and form bound states, namely protons and neutrons. These particles are now non-relativistic and still interact with photons, which makes the baryon diffusion scale drop significantly, so the comoving  size of the matter-antimatter patches freezes after confinement. We can calculate the weighted variance of the baryon asymmetry at the Silk scale $r_0=r_S$ from \eqref{eq:weighed_var}. Note that, after inserting \eqref{eq:nL2_RH} into \eqref{eq:nL2_weighed}, the dependence on the inflation scale cancels out (except for the power spectra, which are nearly scale-invariant), so the Silk scale becomes the only relevant one so that we obtain
\begin{align}
\left.\frac{\langle n_B^2\rangle^{1/2}}{s}\right\vert_{r_S}
\simeq 10^{-36}\left(\dfrac{H_0}{k_0}\right)^{1/2}
\end{align}
for an infrared cutoff $k_0$.

Thus we have the following behaviour for the \textit{rms} baryon asymmetry fluctuations at a given $r_0$ scale
\begin{align}\label{eq:nLs_r0}
\left.\frac{n_B^\mathrm{rms}}{s}\right\vert_{r_0} \simeq
\left\{
\begin{array}{cc}
10^{-36}\left(\dfrac{H_0}{k_0}\right)^{1/2}, \; & r_0<r_S\\
10^{-36}\left(\dfrac{H_0}{k_0}\right)^{1/2}\left(\dfrac{r_S}{r_0}\right)^{7/2},\; & r_0>r_S
\end{array}
\right.
\end{align}
i.e. for regions smaller than the Silk scale, diffusion suppresses baryon fluctuations and the abundance must be obtained through the weighted variance over the Silk length, which does not depend on the inflationary parameters. For larger patches $r_0>r_S$, the averaged abundance is damped as shown in Eq. \eqref{eq:nL2_weighed}.

In conclusion, we see that gravitational leptogenesis effects will induce tiny fluctuations in the baryon asymmetry parameter over regions with a typical size of the Silk length at confinement. Regarding the lepton number asymmetry that is also generated, it is homogenised in a similar manner through free-streaming of electrons and especially neutrinos, which are relativistic until very late stages of cosmic evolution.


\section{Discussion}

In the previous analysis we have considered simple power laws for the scalar and tensor primordial spectra in the whole range of scales with the amplitudes and spectral indices measured from CMB observations. However, as shown in \eqref{eq:nL2_RH}, the produced lepton density depends on the values of the power spectra at the $k_I$ scale,  which can be separated from the scales measured in the CMB by many orders of magnitude. This means that a possible running of the spectral indices could affect the predictions of the model. An interesting possibility
would be the presence of features in the scalar power spectrum at small scales. 
In particular, it has been shown that the presence of (broad) peaks could play an important role in the generation of primordial black holes after inflation \cite{Clesse:2015wea}. Such peaks could in fact enhance in several orders of magnitude the produced lepton asymmetry. 

As shown in  Fig. \ref{fig:ns}, the lepton to entropy ratio strongly depends on the equation of state during the reheating phase and stiff equations enhance the generated asymmetry. This possibility has been recently discussed in \cite{Kamada:2019ewe} where kination dominated reheating scenarios \cite{Spokoiny:1993kt}
have been considered. An interesting consequence of early phases with stiff equation of state is the generation of a  blue tilt in the transfer function of tensor modes \cite{Figueroa:2019paj} which could render the primordial gravitational wave background observable for the sensitivity and  
frequency range of future detectors such as LISA, Einstein Telescope or Cosmic Explorer. Even when considering the \textit{rms} fluctuation at Silk scale, which eliminates the explicit dependency on the inflation scale, both a kination phase and the presence of features in the power spectra introduce a dependency on the particular inflationary scenario.

Even though primordial metric perturbations do not seem to be able to generate
the observed homogeneous asymmetry on Hubble scales, the produced baryon asymmetry could in principle act as a source of baryonic isocurvature perturbations. However, according to the obtained results \eqref{eq:nLs_r0}, these perturbations would be very small for observable scales. Finally, let us mention that beyond the linear regime, gravitational lepton generation in chiral astrophysical systems \cite{delRio:2020cmv} could also
provide potential experimental ways to test the leptogenesis mechanism discussed in this work.

\acknowledgements{We would like to thank Sharma Ramkishor for helpful suggestions. We would also like to thank the Physical Review referees for their insightful comments. This work has been supported by the MINECO (Spain) project PID2019-107394GB-I00 (AEI/FEDER, UE). A.D.M. acknowledges financial support by the MICIU (Spain) through a Formación de Profesorado Universitario (FPU) fellowship FPU18/04599.}

\bibliography{metric_leptogenesis}

\end{document}